\documentclass[twoside,twocolumn,9pt]{article}
\usepackage{extsizes}                         
\usepackage[left=1.5cm,right=1.5cm,top=1.785cm,bottom=2.0cm]{geometry}
\usepackage[T1]{fontenc}
\usepackage{mathptmx}                          
\usepackage{droidsans}                         
\usepackage[super,sort&compress,comma]{natbib} 
\usepackage{graphicx}
\usepackage{amsmath,amssymb,bm}
\usepackage{siunitx}
\usepackage[usenames,dvipsnames]{xcolor}
\usepackage{balance}                           
\usepackage[format=plain,justification=justified,singlelinecheck=false,
            font={stretch=1.125,small,sf},labelfont=bf,labelsep=space]{caption}
\usepackage[compact]{titlesec}

\usepackage{microtype}      
\titleformat*{\section}{\normalsize\bfseries\sffamily\raggedright}
\titleformat*{\subsection}{\small\bfseries\sffamily\raggedright}
\titleformat*{\subsubsection}{\small\itshape\sffamily\raggedright}
\usepackage{hyperref}
\hypersetup{colorlinks=true,linkcolor=blue,citecolor=blue,urlcolor=blue}
\begin{document}

\twocolumn[{%
  \begin{@twocolumnfalse}
  \begin{center}
    {\LARGE\bfseries\sffamily Size matters more than packing in bimodal colloidal gel compositions\par}
    \vspace{1.5ex}
    {\large Robert A. Campbell\textsuperscript{1}, Ziye Zhuang\textsuperscript{2},
            Ali Mohraz\textsuperscript{2}, Safa Jamali \textsuperscript{1}\par}
    \vspace{1ex}
    {\itshape\small
      \textsuperscript{1}Department of Mechanical and Industrial Engineering,
        Northeastern University, Boston, MA 02115, USA\\
      \textsuperscript{2}Department of Chemical and Biomolecular Engineering,
        University of California, Irvine, CA 92697, USA\par}
  \end{center}
  \vspace{2ex}
  \vspace{1.5ex}
  {\noindent\textbf{Abstract}\quad Colloidal gels are frequently modeled as monodisperse particle networks, although practical formulations commonly contain particles with multiple characteristic sizes. Here, we use large-scale, hydrodynamically resolved simulations of colloidal depletion gels to isolate the effects of particle size and local packing in bimodal systems with a small-to-large size ratio of 1:2. Increasing the large-particle fraction introduces new heterotypic angular motifs and substantially increases the fraction of bonds participating in tetrahedral structures, with a maximum at intermediate composition. However, these additional rigid motifs do not reorganize into larger or more highly connected tetrahedral aggregates. The mean coordination and characteristic aggregate size remain nearly composition independent. By contrast, the void and cluster-size distributions coarsen systematically as the large-particle fraction increases. These mesoscale distributions largely collapse when normalized by a composition-dependent particle length scale, indicating that changes in composition primarily rescale gel architecture rather than producing distinct rigid-network topologies. An elastic modulus estimated using Cauchy–Born theory similarly follows this effective length scale more closely than the abundance of local tetrahedral motifs. These results show that, for moderate size disparity, particle size controls the structural scale and predicted mechanical response of bimodal colloidal gels more strongly than enhanced local packing.\par}
  \vspace{1.5ex}
  \vspace{3ex}
  \end{@twocolumnfalse}
}]


\section*{Introduction}
Colloidal gels are space-spanning particulate networks formed when attractive interactions cause the aggregation, and eventual percolation, of suspended colloid particles\cite{Zaccarelli2007,Lu2008,Royall2021}. Their connectivity allows these systems to sustain stress and exhibit viscoelastic solid-like behavior at low particle volume fraction\cite{ZacconeWuDelGado2009,Bantawa2023,Leocmach2014,Lu2013,Trappe2001}, making them central to the production of foods, coatings, personal-care products, pharmaceuticals, and many other formulated materials. Because their mechanical response emerges from the architecture of the particle network, colloidal gels provide a direct link between formulation variables—including particle size, volume fraction, attraction strength, and interaction range—and macroscopic rheology.

Most classical descriptions of colloidal-gel structure and mechanics have been developed for monodisperse particles\cite{Lin1989,Krall1998,Shih1990}. These models have provided important relationships between local coordination, cluster structure, characteristic length scales, and elastic response in uniform populations of particles, regardless of their exact size. However, industrial and naturally occurring suspensions are rarely monodisperse. Even moderate size dispersity within a system can alter aggregation kinetics, local packing, and the hierarchical organization of the load-bearing network. Recent studies indicate that particle-size dispersity can reorganize local structural motifs even when the different particle populations remain well mixed. Waheibi and Hsiao observed a re-entrant change in local disorder in bidisperse colloidal systems, with the largest departure from monodisperse packing occurring at intermediate size ratios \cite{WaheibiHsiao2024}. They associated this behavior with changes in the formation and organization of rigid tetrahedral motifs. Similarly, Lonial and Weeks \cite{LonialWeeks2026} found that highly polydisperse gels assemble through approximately random particle pairing, without strong size-selective segregation. Nevertheless, larger particles became preferentially incorporated into locally rigid tetrahedral environments as the network evolved. Together, these findings demonstrate the extent to which size disparity can influence the distribution of locally rigid structures in colloidal gels. A central unresolved question is whether these effects are composition-dependent. In comparison with established monodisperse models, where network features can be described in terms of a system-wide particle scale, it is unclear if the behavior of bimodal gels is governed primarily by composition-dependent changes in local packing or by an effective particle-size scale introduced by the mixture.

These observations motivate a hierarchical view of colloidal-gel mechanics. At the particle scale, compact motifs such as tetrahedra provide locally rigid building blocks \cite{Tsurusawa2023,Zhang2019,Dias2025,Koeze2018}. The spatial arrangement and connectivity of these motifs generate mesoscale strands, junctions, clusters, and voids, which collectively determine how stress is transmitted across the percolated network \cite{Bantawa2023,Bouzid2017,Kim2014,Whitaker2019,Zia2014}. Local structural descriptors alone are therefore generally insufficient to predict the bulk response: gels with comparable coordination numbers can exhibit substantially different elastic moduli because similar local units may be organized into very different network-scale architectures \cite{Rocklin2021}. The mechanically relevant question is not only how many rigid motifs form, but whether those motifs connect into an extended backbone and how that backbone sets the characteristic structural length scale of the gel. For instance, changes in local structure can alter how these hierarchical features emerge from the network, and influence both gelation dynamics and the bulk mechanics of an otherwise equivalent system \cite{Zhuang2026}. 

Mesoscale theories of colloidal-gel elasticity formalize this hierarchical picture through a characteristic structural length scale. Fractal-floc scaling relates the elastic modulus to cluster size, cluster compactness, and the connectivity between clusters \cite{Shih1990,deRooij1994,Krall1998}, while Cauchy–Born approaches connect network elasticity to the deformation of mesoscale structural units and their interactions \cite{ZacconeWuDelGado2009,Whitaker2019,Nabizadeh2024}. In the Cauchy-born framework, cluster size and cluster-cluster interactions have been further demonstrated to be a predictive tool for scaling of the bulk elastic modulus \cite{ZacconeWuDelGado2009,Whitaker2019,Nabizadeh2024}. The void structure provides a complementary measure of the network mesh size and can become particularly important when particle inclusions, size dispersity, or other sources of heterogeneity distort the gel architecture \cite{Li2023}.

Here, we use large-scale, hydrodynamically resolved simulations to separate the effects of particle size and local packing in bidisperse depletion gels with a radius ratio $a_{L}/a_{S}=2$. We hold the total colloid volume fraction and depletion condition fixed while varying the large-particle volume fraction $x_L=\phi_{L}/\phi$ from 0 to 1. We quantify composition-dependent changes in angular motifs, tetrahedral participation, tetrahedral-aggregate topology, coordination, void structure, and cluster size, and then compare these structural measures with an elastic modulus estimated using Cauchy–Born theory. The results reveal a clear separation between structural scales. Bimodal composition strongly modifies local packing and tetrahedral abundance but only weakly changes the connectivity and characteristic extent of the rigid aggregates. The dominant effect is instead a systematic coarsening of clusters and voids that can be described using a composition-dependent effective particle size. Thus, for the moderate size disparity considered here, particle size sets the network scale and predicted elastic response more strongly than local packing efficiency.

\section*{Methods}

\subsection*{Bimodal depletion interactions}
We use a size-weighted Morse potential to accurately mimic size-dependent effects in bimodal depletion systems. In the Asakura-Oosawa-Vrij model of depletion \cite{Asakura1958,Vrij1976} the attraction strength is set by the volumetric exclusion of a small polymer depletant. A large colloid radius therefore increases the overlap volume, increasing its effective attraction as compared to a smaller sized colloid. In this case, the same depletant concentration and depletant size will produce heterogeneous interactions between small-small, small-large, and large-large colloid pairs. We do not simulate explicit depletant particles in our systems. Instead, we include these size effects as a size-weighted Morse potential. We scale the attraction minimum with the average size of each interacting colloid pair, such that $U_0^{ij}=D_0(a_{i}+a_{j})/2$. At particle size ratio 1:2, this approximation produces attraction strengths in agreement with the Asakura-Oosawa-Vrij model. Pair-wise interactions are therefore calculated as:

\begin{equation}
    \mathbf{U}_{ij}^{Morse} (r_{ij}) = U^{ij}_0 \big(e^{-2\alpha (r_{ij}-(a_i+a_j))}- 2e^{-\alpha (r_{ij}-(a_i+a_j))}\big)
    \label{eq:morse}
\end{equation}
\\
\noindent
where $a_i$ and $a_j$ are the particle radii, $r_{ij}$ is the center-center separation distance, $kT$ is the system temperature, and $\alpha$ is the attraction range parameter. In this study all systems are simulated at a fixed attraction range representing the same depletant, with $\Delta = a_S/r_g = 0.05$, simulated as $\alpha/a_S=60$. All systems are simulated at $D_0 = 12kT$. This represents the effective interaction between two small-small particles. In these bimodal systems, there is always a corresponding attraction hierarchy, such that $U_0^{SS}=D_0$, $U_0^{SL}=1.5D_0$ and $U_0^{LL}=2D_0$. 

\subsection*{Simulation}
Colloidal systems were simulated using a Core-Modified Dissipative Particle Dynamics (CM-DPD) method \cite{Whittle2010,Jamali2015,Boromand2017} implemented in a custom version of the open-source molecular dynamics toolkit HOOMD-blue \cite{Anderson2020}. CM-DPD is a discrete fluid model where both colloids and solvent are included explicitly as particles. The addition of solvent particles ensures that long-range hydrodynamics are accurately transmitted between colloids. Integration is performed at constant number of particles, constant volume, and constant temperature (NVT), using a two-step velocity-Verlet algorithm. Throughout the simulation, the system temperature is maintained at $kT=0.1$ through a fluctuation-dissipation relation. The equation of motion for particle $i$ with mass $m_i$ and velocity $\mathbf{v}_i$ is described by six pair-wise forces:
\begin{equation}
m_{i} \frac{d\mathbf{v}_i}{dt} = \sum_{i,i \neq j}^{N}\big( \mathbf{F}_{ij}^{Cons} + \mathbf{F}_{ij}^D + \mathbf{F}_{ij}^R + \mathbf{F}_{ij}^{M} + \mathbf{F}_{ij}^H + \mathbf{F}_{ij}^{Cont}\big)
\label{eq:DPD-colloid}
\end{equation}
Solvent-solvent and solvent-colloid interactions are described by the first three force terms, with all other forces set to zero. These three terms represent a soft conservative force, $F_{ij}^{C}$ , a dissipative force, $F_{ij}^D$, and a random force, $F_{ij}^R$, defined as:
\begin{equation}
\mathbf{F}_{ij}^{Cons} = a_{ij} \omega_{ij}(r_{ij}) \mathbf{e}_{ij}
\label{eq:DPD-Conservative}
\end{equation}
\begin{equation}
\mathbf{F}_{ij}^D = -\gamma_{ij} \big[\omega_{ij}(r_{ij})\big]^2 \big( \mathbf{v}_{ij} \cdot \mathbf{e}_{ij} \big) \mathbf{e}_{ij}
\label{eq:DPD-Dissipative}
\end{equation}
\begin{equation}
\mathbf{F}_{ij}^R = \sigma_{ij} \omega_{ij}(r_{ij}) \Theta_{ij} \Delta t^{-2}\mathbf{e}_{ij}
\label{eq:DPD-Random}
\end{equation}
\noindent
The controlling parameter for the conservative force is the maximum attraction strength for the background fluid, $a_{ij}=25kT$, as calculated from the compressibility and density of water. The $\mathbf{e}_{ij}$ parameter is the directional vector of the contact, and the pair-specific weight term, $\omega_{ij}(r_{ij})=(1-(r_{ij}/r_{cut}))$, is calculated from the separation distance $r_{ij}$ and a maximum neighbor cut-off distance, $r_{cut}$. To satisfy the fluctuation-dissipation theorem, the controlling parameters for dissipative and random forces are defined in terms of each other, such that $\sigma_{ij}=\sqrt{2\gamma_{ij}kT}$. Additionally, the viscous resistance parameter, $\gamma_{ij}$, scales with the background viscosity, $\eta_0$, here set as $\gamma_{ij}=4.5, \eta_0=0.3$. The theta parameter, $\Theta_{ij}$, is a random-number white noise parameter, and $\Delta t$ is the integration timestep size. 

In colloid-colloid interactions, we set $F_{ij}^{C}=0$ and the interactions are defined by the last five force terms. This includes the dissipative and random forces as well as the Morse attractive force $F_{ij}^{M}$, a short-range lubrication term, $\mathbf{F}_{ij}^{H}$, and a hard-sphere repulsive contact force, $\mathbf{F}_{ij}^{Cont}$. These forces are defined in terms of the surface-surface particle separation distance, $h_{ij}=r_{ij}-(a_i+a_j)$, as:
\begin{equation}
\mathbf{F}_{ij}^{M} = \frac{d\mathbf{U_{ij}^M}}{d\mathbf{r}_{ij}} = -2 \alpha U_0^{ij} \big( e^{-2 \alpha h_{ij}} - e^{- \alpha h_{ij}}\big)
\label{eq:DPD-MorseForce}
\end{equation}
\begin{equation}
\mathbf{F}_{ij}^{Cont} = 
    \begin{cases}
      f_{cont}(1+h_{ij}) &  h_{ij} \leq 0,\\
      f_{cont}(1-\frac{h_{ij}}{\delta_C}) &  0 \leq h_{ij} \leq \delta_C,\\
      0 &  h_{ij} > \delta_C,
    \end{cases}  
\label{eq:DPD-Contact}
\end{equation}
\begin{equation}
\mathbf{F}_{ij}^H = \mathbf{F}_{ij}^{sq} = -\mu_{sq}\big( \mathbf{v}_{ij} \cdot \mathbf{e}_{ij} \big)\mathbf{e}_ij
\label{eq:DPD-Hydro}
\end{equation}
\noindent
The contact force is defined by a magnitude $f_{cont}/kT=100$ with an activation range $\delta_C=0.001$ from the particle surface. The hydrodynamic lubrication force is calculated from the squeezing-coefficient for two particles interacting as a coordinated dumbell at low Reynolds number\cite{BallMelrose1997}. This is the pair-drag term $\mu_{ij}=3\pi\eta_0(4a_ia_j/(a_i+a_j))^2/(8h_{ij})$. Small colloids are assigned a radius $a_S=1$, and large colloid are assigned a radius $a_L=2$. Solvents are represented as point particles distributed with a number density $\rho=3$. Densities are matched between particles by assigning the unit mass to solvents, $m_S=1.0$, and setting the mass of each colloid as $m_C = 4/3 \rho \pi a_C^3$. 

To ensure comparable structural features across all simulations, all systems are initialized in a cubic box of side length $L= 70a_S$ with periodic boundary conditions. The total volume fraction of colloids is held constant at $\phi = 0.2$. The bimodal systems are divided into two sub-populations by volume, such that each system can be defined by the large particle fraction $x_L=\phi_L/\phi=[0.00, 0.25, 0.50, 0.75, 1.00]$. This produces a variable number of particles between each system: $x_L=0.0$ has 16,377 colloids; $x_L=0.25$ has 12,795 colloids (12,283 small, 512 large); $x_L=0.50$ has 9,213 colloids (8,189 small, 1,024 large); $x_L=0.75$ has 5,629 colloids (4,094 small, 1,535 large); and $x_L=1.0$ has 2,047 colloids. Because $\phi$ and $\rho$ are both constant, all systems contain an additional 823,200 solvent particles to carry the long-range hydrodynamic interactions.

The simulations were performed in two steps. In the first step, all particles were randomly placed in the simulation box and allowed to briefly equilibrate without colloid-colloid attraction. This allows the simulation to resolve non-physical overlaps and fully distribute particles under quiescent conditions. In the second step, attractive interactions were added and the system was allowed to aggregate. The formation of a quasi-steady state network was verified by monitoring structural assembly of a space-spanning percolated network, as defined by high average coordination number $\langle Z \rangle$, more than 80\% of the system participating in the largest connected component (LCC), percolation across the entire span of the system box, and a sub-diffusive mean squared displacement, in accordance with previous methods \cite{Mangal2024,Nabizadeh2024,PanizHaghighi2025}.

\begin{figure*}[ht!]  
\centering
  \includegraphics[height=8cm]{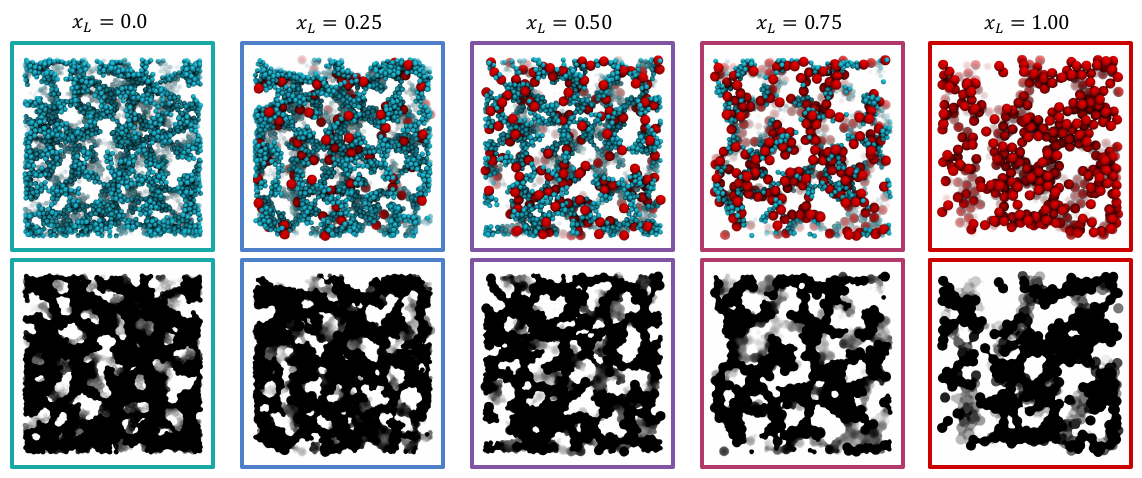}
  \caption{Change in bimodal composition alters colloidal gel structure. Visualization of a representative $38a_S \times 38a_S \times 13a_S$ size slab of simulated colloidal gels at total volume fraction $\phi=0.2$ and varying large particle fractions $x_L=[0.0, 0.25, 0.5, 0.75, 1.0]$. The top row shows the system color-coded by particle size, with small particles in teal and large particles in red. The bottom row shows the surface projection of the same z-slice, where individual particles are merged together into a single black network surface. In both rows, the color of the bounding box signifies the bimodal composition $x_L$. Left to right, these are: 0.0 (teal), 0.25 (blue), 0.5 (purple), 0.75 (maroon), and 1.0 (red).}
  \label{fig:comp}
\end{figure*}

\subsection*{Network analysis}
A contact network is constructed from the final simulation state of each system using the set of colloidal pairs that fall within the interaction limit $h_{ij}<\Delta$. We label each bond by it's component particles (small-small, small-large, or large-large) and use this network to calculate the angle distribution for each system. Angles between incident edges are calculated from their unit length as $\theta_{ijk}=\arccos ( \hat{\mathbf{e}}_{ij} \cdot \hat{\mathbf{e}}_{ik})$ and converted to degrees. These angles are then labeled by their component triplets as small-small-small (SSS), small-small-large (SSL), small-large-small (SLS), large-small-large (LSL), large-large-small (LLS), or large-large-large (LLL).

Additionally, we label all particles and bonds that participate in rigid tetrahedral units within the network. A single rigid tetrahedra is defined as a 4-clique, a set of six bonds that connect four adjacent particles. This structure has been established as a simple and useful unit of local rigidity in colloidal gels \cite{LonialWeeks2026}. We also classify sets of seven fully connected colloids (7 nodes, 16 edges) as pentagonal bipyramids, a stricter definition of rigid units that can distinguish between amorphous and crystalline pathways to rigidity percolation \cite{Tsurusawa2023}. However, we do not observe the formation of any pentagonal bipyramids in our gels, consistent with recent experiments that emphasized the role of simple tetrahedra in the low-$\phi$ gel regime \cite{WaheibiHsiao2024,LonialWeeks2026}. 

\subsection*{Characterization of additional multiscale structures}

We also identify the mesoscale cluster structure of the network using an established graph-based Gaussian Mixture Model (GMM) method \cite{Nabizadeh2024,Zhuang2026}. The contact network is projected into a high-dimensional latent space using the semi-supervised node2vec model \cite{Grover2016node2vec}, in which a stochastic gradient descent approach is used to learn the feature representation of the network. This high-dimensional system is then reduced using Uniform Manifold Approximation and Projection (UMAP), producing a lower-dimensional representation of the network that is weighted by the relationships between regions of connected neighbors \cite{Mcinnes2020umap,McInnes2018}. A GMM is then used to group the original particles into clusters based on the topology of this latent space. This method assumes that the network can be described as an unknown multivariate Gaussian distribution. It uses an expectation-maximization algorithm to solve for $k$ distributions that describe the total, multivariate system\cite{Dempster1977}. We identify the optimal number of clusters by running GMM across a wide range of $k$ values and selecting the one that maximizes the Bayesian Information Criterion (BIC).

After assigning all particles to unique clusters, we validate the physical identity of these clusters in the original 3D space. In this study, all of our original networks were fully connected (all particles belong to a single largest connected component), so we expect each cluster to also be a fully-connected subgraph. However, since GMM is performed in topological latent space and does not use physical particle position, it is possible that two topologically similar regions of the network that are far apart from each other could be grouped into a single cluster. In practice this is not common in our systems, but we still implement a physicality check. If a cluster subgraph is not fully connected, we evaluate the size of it's disconnected components. If the largest component makes up more than $90$\% of the cluster, we reassign the remaining disconnected components to their nearest physical neighbor. Otherwise we split the cluster into it's constituent components. If these components are more than $10$\% of the original cluster size, we define them as new clusters, otherwise we reassign them to their nearest neighboring physically connected cluster. 

\section*{Results and discussion}

We organize the results in three hierarchical levels, to distinguish changes in local packing from changes in the characteristic network scale. We first establish how bimodal composition changes the overall gel morphology and the geometry of particle contacts. We then determine whether the resulting increase in compact local motifs produces larger or more highly connected rigid aggregates. Finally, we quantify the composition-dependent cluster and void scales and compare their influence on the predicted elastic response. This progression allows us to identify not merely whether bimodality changes the gel structure, but which level of the structural hierarchy controls its mechanics.

\begin{figure*}[t] 
\centering
  \includegraphics[height=11cm]{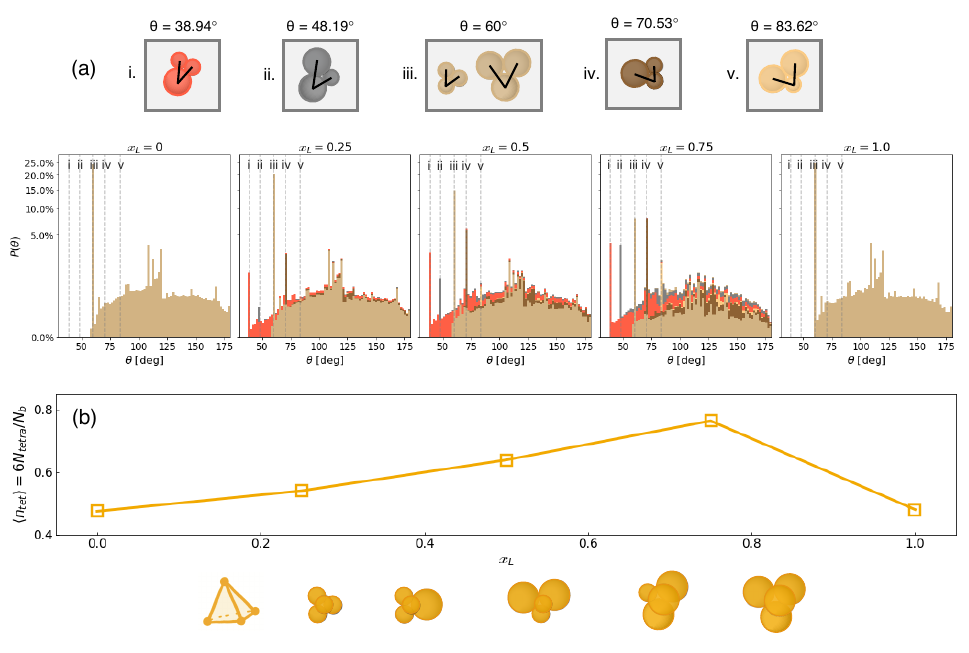}
  \caption{Changing bimodal composition alters particle packing. (a) The distribution of three-body angles in each system by bimodal structure type. The top row provides a schematic representation of the five unique 3-body bimodal structures that can form in these systems and their associated close-packed angle, arranged from smallest to largest: (i) small-large-small (SLS, $\theta=38.94${\textdegree}, dark orange) (ii) small-large-large (SLL, $\theta=48.19${\textdegree}, dark grey) (iii) small-small-small or large-large-large (SSS and LLL, $\theta=60${\textdegree}, dark tan) (iv) large-large-small (LLS, $\theta=70.53${\textdegree}, brown) and large-small-large (LSL, $\theta=83.62${\textdegree}, light tan). The second row shows the distribution of angles in each composition ($x_L$=[0.0, 0.25, 0.5, 0.75, 1.0]). The total distribution is divided by angle type and stacked to display the fraction of total angles that are found with a given $\theta$ value, and types of bimodal structures that compose those angle in the network. (b) The fraction of bonds that participate in tetrahedra in each system, $\langle n_{tet} \rangle = 6N_{tetra}/N_b$. The bottom row provides a schematic representation of a tetrahedral unit, and the different types of bimodal and monodisperse tetrahedra that can appear in these systems.}
  \label{fig:packing}
\end{figure*}

\subsection*{Bimodal composition alters colloidal gel structure}
All five compositions form a single, compositionally mixed, percolated network under constant total particle volume fraction and high attraction strength. Figure \ref{fig:comp} shows a representative slab of dimensions $38a_S \times 38a_S \times 13a_S$ of each gel. The top row of images shows each system color coded by particle size (small particles in teal, large particles in red). Across the three bimodal compositions we see no evidence of phase-separation into population-specific networks. Instead, both small and large particles appear fully dispersed throughout the same LCC. This is true even at low ($x_L = 0.25$) and high ($x_L = 0.75$) large-particle fractions, where large differences in the number fraction of small and large particles does allows large, monodisperse regions to form; however, these regions are notably still bridged by heterotypic contacts. Additionally, the network structure appears to coarsen as the large-particle fraction increases, producing larger voids between strands. The surface projections in the bottom row of Fig.\ref{fig:comp} make this mesoscale coarsening more apparent: the characteristic strand and void dimensions increase with $x_{L}$, even though the total particle volume fraction remains fixed. At $a_L/a_S=2$, one large particle occupies the same solid volume as eight small particles. Increasing $x_{L}$ therefore decreases the number of particles required to occupy a fixed solid volume and introduces a larger geometric unit from which the gel network is constructed. This effect alone should increase the characteristic dimensions of strands, clusters, and voids. At the same time, heterotypic contacts create angular configurations that are unavailable to monodisperse particles and may promote more compact local packing. These two effects lead to competing mechanical expectations. Geometric coarsening should increase the network mesh size and reduce the number of parallel load-bearing pathways, whereas enhanced local packing could increase the abundance and connectivity of rigid motifs. We therefore first quantify the local packing response and then determine whether it propagates into the topology of the mechanically rigid network.

\begin{figure}[ht!] 
\centering
  \includegraphics[height=4.5cm]{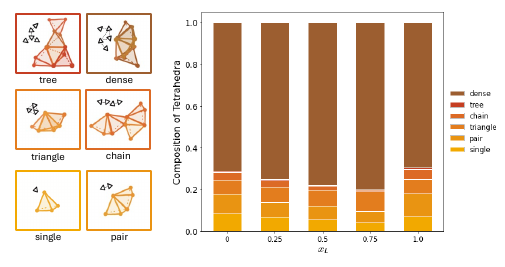}
  \caption{Composition of tetrahedral aggregates across all five bimodal fractions. The left schematics shows the six classes of tetrahedral aggregates that are measured in these systems: single tetrahedra (bottom left, yellow), pairs of tetrahedra (bottom right, gold), fully-connected triangles of three tetrahedra (center left, orange), chains of three or more tetrahedra (center right, dark orange), branching tree structures (top left, red), and dense aggregates of four or more tetrahedra that are not fully connected (top right, brown). The graph on the right shows the relative composition of these six classes across the five bimodal fractions, $x_L=[0,0.25,0.5,0.75,1.0]$.}
  \label{fig:tetra}
\end{figure}

\begin{figure}[ht!] 
\centering
  \includegraphics[height=4cm]{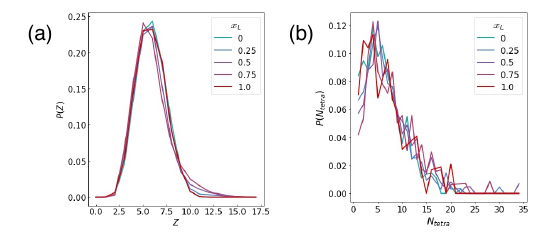}
  \caption{Distribution of local contacts and tetrahedral aggregates. (a) The coordination number distribution for colloidal bonds across all five large particle fractions, $x_L$: 0.0 (teal), 0.25 (blue), 0.05 (purple), 0.75 (maroon), and 1.0 (red). (b) The distribution of tetrahedral aggregates sorted by number of individual tetrahedral units, ${N_{tetra}}$, across the same five system compositions.}
  \label{fig:meanfield}
\end{figure}

\subsection*{Bimodality enhances local packing and tetrahedra \\participation}
Having established that the network coarsens at the mesoscale, we first ask whether the same change in composition also produces a denser and more locally rigid particle arrangement. Figure \ref{fig:packing}a, verifies that a change in microscale organization is taking place. Examining the distribution of angles in the network, we see that the amorphous nature of colloidal gel structure allows for a wide distribution of 3-body angles to form across the system; however, there are distinct peaks corresponding to the most preferred structural motifs. In both monodisperse systems we see significant peaks at $\theta=60${\textdegree} corresponding to an energy-minimizing densely packed triplet, with secondary peaks near $\theta=120${\textdegree} and $\theta=180${\textdegree} where these triplets cluster together. The $\theta=60${\textdegree} peak remains prevalent in all bimodal systems; however, the secondary peaks decrease as alternate types of heterotypic triplets become more common. Labeling each triplet by its particle composition confirms that this change in network structure is the result of heterotypic packing. At $x_L=0.25$ we see new peaks emerge as small particles begin clustering around their large neighbors at: $\theta=38.94${\textdegree} (SLS triplets), $\theta=48.19${\textdegree} (LSL triplets), and $\theta=70.53${\textdegree} (LSS triplets). The magnitude of these peaks continues to increase as the fraction of large particles increases and the formation of heterotypic triplets become more likely. At $x_L=0.5$ a peak at $\theta=83.62${\textdegree} (LSL triplets) also emerges, now that the number of large particles has increased enough for them to preferentially cluster around small particles, and at $x_L=0.75$ all four of these bimodal triplets approach a comparable likelihood to $\theta=60${\textdegree}. Although there are still small peaks near $\theta=120${\textdegree} and $\theta=180${\textdegree}, these structures are now statistically comparable to other, more disordered structures. Figure \ref{fig:packing}b shows that the change in angle distribution corresponds to an increase in the formation of mechanically rigid tetrahedral clusters of 4 particles connected by 6 bonds. These serve as a minimal rigid units for central-force dominated systems like these, where each bond restricts a single degree of freedom. Comparing the average fraction of bonds that participate in tetrahedra $\langle n_{tetra} \rangle$ across the five systems, we see that it increases from around 40\% in monodisperse systems to 80\% when $x_L=0.75$. Therefore, a high fraction of large particles (where the number fraction of small and large particles is nearly equivalent) maximizes the formation of individual tetrahedral units. The nonmonotonic tetrahedral response isolates a distinctly packing-driven effect. The $x_L=0.75$ system contains the largest proportion of bonds associated with tetrahedral motifs, even though it does not have the largest particle or network length scale. This composition therefore provides a useful test of whether enhanced local packing is sufficient to reorganize the load-bearing structure and control the mechanical response.

An increase in tetrahedral participation can influence network mechanics only if the additional rigid units organize into aggregates that are capable of transmitting stress beyond the scale of a single motif. We therefore classify the topology of connected tetrahedral aggregates and determine whether the composition that maximizes tetrahedral abundance also produces more extended or more highly connected rigid structures. Figure \ref{fig:tetra} compares the relative composition of seven different topologies of tetrahedral aggregates in each system: single tetrahedra, pairs of tetrahedra that share at least one edge or face, triangles of three fully connected tetrahedra, chains of three or more tetrahedra with a maximum coordination number of $Z_{tetra}^{max}=2$, junctions or larger branching trees with four or more tetrahedra and $Z_{tetra}^{max}\geq3$, dense aggregates of four or more tetrahedra. We also track the number of fully connected pentagonal bipyramids, units of five tetrahedra arranged around a common edge, forming a compact seven-particle structure with local fivefold symmetry. Although we expect this topology to be rare at low particle volume fraction, we treat this structure separately because it represents a particularly dense and locally favored arrangement, where its fivefold symmetry prevents periodic space filling. Its abundance therefore provides a measure of local densification and geometric frustration rather than crystalline ordering or network-spanning rigidity; however, we do not measure the formation of any true pentagonal bipyramids, consistent with recent experimental results\cite{LonialWeeks2026,WaheibiHsiao2024}. Across compositions, we instead find that the system is always dominated by more loosely connected dense assemblies. Additionally, as $x_L$ increases, the fraction of tetrahedra that form pairs, chains, and branches decreases in favor of these denser topologies. The increase in tetrahedral participation therefore reflects a redistribution toward already compact local assemblies rather than the formation of longer tetrahedral chains or a percolated tetrahedral domain. In particular, the reduction in pair and chain populations at larger $x_L$ suggests that newly formed tetrahedra are incorporated locally into dense aggregates rather than extending the rigid structure over larger distances. The additional local rigidity is thus topologically localized.

The topology analysis indicates that bimodality changes how tetrahedra are locally grouped but does not produce an extended tetrahedral backbone. We next test whether this apparent invariance is also reflected in ensemble-averaged measures of local connectivity and aggregate size. Despite the increased formation of dense aggregates we do not see major restructuring of tetrahedral structures in the total system. Figure \ref{fig:meanfield}a shows that, surprisingly, local structure remains comparable across all five compositions. Despite the ability to form high-coordination SL contacts in bimodal systems, the total coordination number distribution remains centered around $\langle Z \rangle \approx 6$ at all values of $x_L$. Similarly, although the bimodal systems do produce larger tetrahedral aggregates these remain quite rare, and the total distribution of tetrahedra also does not change. Figure \ref{fig:meanfield}b shows that the monodisperse systems already produce a few tetrahedral aggregates made of as many as 20 individual tetrahedra. We find that bimodal systems can produce aggregates with as many as 34 individual units, but the average size of tetrahedral aggregates does not meaningfully change. Instead, it appears that, at least at this moderate size ratio of 1:2, system-wide microstructure is set more by the total particle volume fraction than either tetrahedra formation or particle stoichiometry.

\begin{figure*}[ht!]
 \centering
 \includegraphics[height=10cm]{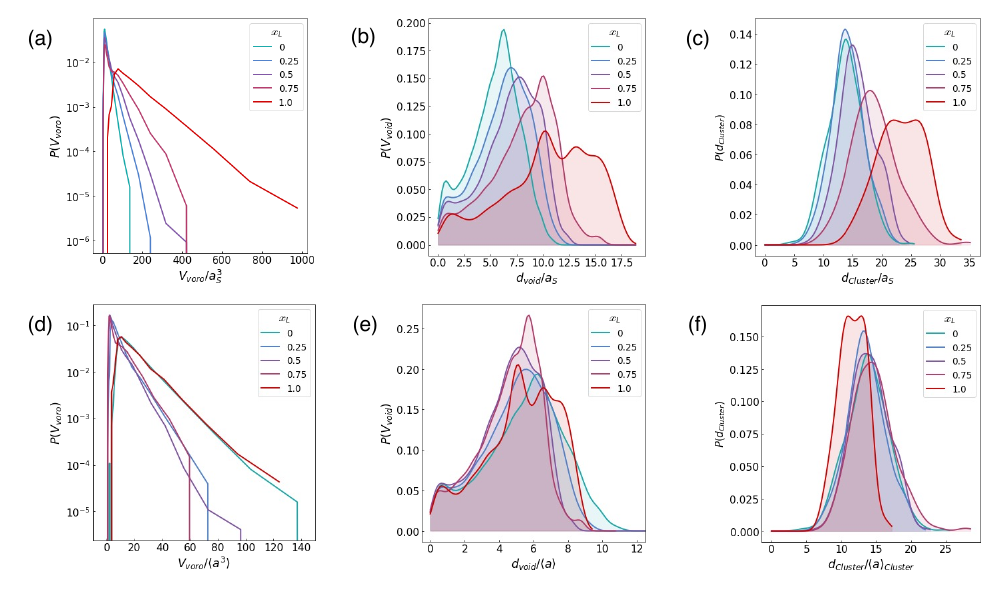}
 \caption{Distribution of multiscale structures across bimodal compositions. All plots include data for five large particle volume fractions, $x_L$: 0.0 (teal), 0.25 (blue), 0.05 (purple), 0.75 (maroon), and 1.0 (red). (a) The Voronoi volume distribution in terms of the simulation unit length, $a_S^3$. (b) The void size distribution by void diameter, $d_{void}$, reported in terms of the simulation unit length, $a_S$. (c) The cluster size distribution in terms of the diameter of the sphere that encloses each cluster $d_{cluster}$, reported in terms of the simulation unit length $a_S$. (d) The same Voronoi volume data, rescaled in terms of the average volumetric unit of colloid size in each system, $\langle a^3 \rangle$. (e) The same void size data, rescaled in terms of the average colloid size in each system, $\langle a \rangle$. (f) The same cluster size data, rescaled per cluster in terms of the average colloid size of it's component particles $\langle a \rangle_{Cluster}$.}
 \label{fig:coarsening}
\end{figure*}

Taken together, Figs. \ref{fig:packing}-\ref{fig:meanfield} establish an important separation between local packing and network-scale organization. Bimodality changes the probability of forming compact rigid motifs, but it does not substantially alter the mean coordination, the characteristic tetrahedral-aggregate size, or the extent of the tetrahedral network. The pronounced morphological coarsening observed in Fig. \ref{fig:comp} must therefore originate at a higher level of the structural hierarchy. We next quantify that mesoscale change using particle-centered Voronoi volumes, void dimensions, and cluster sizes.

\subsection*{Composition sets the mesoscale structural length}
Having shown that local rigidity is only weakly reorganized by composition, we now ask whether the dominant effect of bimodality is instead to change the characteristic length scale of the network. Figure \ref{fig:coarsening}a shows that the distribution of Voronoi volume accessible to each particle becomes significantly more heterogeneous as the fraction of large particles increases. This increase in variance corresponds to the change in volume distribution within the network structures, as large particles effectively replace small-small clusters in the gel strands. This corresponds to a monotonic increase in void size seen in Figure \ref{fig:coarsening}b. As the $x_L$ increases, the fraction of very small voids, which exist within clusters, decreases and the total distribution widens to peak at a larger average diameter. Interestingly, this coarsening also shows up in the cluster size distribution, shown in Figure \ref{fig:coarsening}c. As large particles make up more of the system, large-large clusters become more common and the average cluster size increases.

Figures \ref{fig:coarsening}d,e, and f reveal that these changes can be summarized as a function of the effective length scale of each system. Instead of using the same length scale for all systems (simulation unit length, equivalent to the small colloid radius $a_S=1$), we normalize the Voronoi volume distribution and the void size distribution by the average particle size in each system, and we normalize each cluster size by the average particle size within that cluster. Rescaling our analyses in this regard collapses each distributions onto a single curve. The three bimodal compositions now collapse onto a single, narrower Voronoi volume distribution (\ref{fig:coarsening}d), corresponding to the increased capacity for local packing. We also see some distortion and broadening of the peak for void size (\ref{fig:coarsening}e) and cluster size (\ref{fig:coarsening}f) which we primarily attribute to statistical error from the fewer number of particles present in the $x_L=1.0$ data; however, the overall trend remains clear. Bimodal composition effectively alters the relative length scale of mesoscale features in a network.

\begin{figure}[ht]
 \centering
 \includegraphics[height=5.5cm]{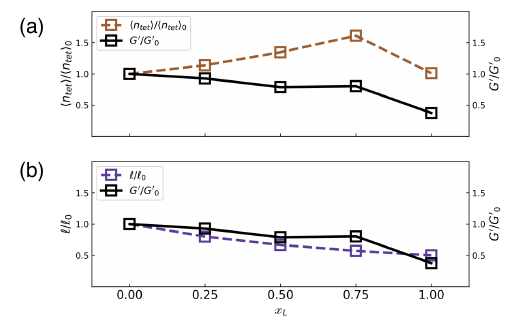}
 \caption{Comparison of structural and mechanical scaling across all five large particle fractions. (a) Comparison between the fraction of bonds that participate in tetrahedra $\langle n_{tetra} \rangle$, normalized by the fraction of bonds that participate in tetrahedra in the $x_L=0.0$ system $\langle n_{tetra} \rangle_0$, (brown) and the Cauchy-Born prediction of bulk elasticity $G'$, normalized by the prediction for the $x_L=0.0$ system $G'_0$ (black). (b) Comparison between the effective particle length scale of the system, as calculate from the average particle size, $\ell=1/\langle a \rangle$, normalized by the length scale of the $x_L=0.0$ system $\ell_0$ (purple) and the same normalized Cauchy-Born prediction of bulk elasticity, $G'/G'_0$.}
 \label{fig:mech}
\end{figure}

This structural collapse provides a direct means of testing which descriptor more closely controls the mechanical response. We therefore compare the composition dependence of the Cauchy–Born elastic modulus with two quantities: tetrahedral participation, which measures enhanced local packing, and the characteristic mesoscale length, which measures network coarsening. The modulus exhibits a modest nonmonotonic enhancement near $x_L=0.75$ , where tetrahedral participation is maximized (Fig. \ref{fig:mech}a). This indicates that local packing does provides a measurable contribution to the multi-scale features that determine bulk rigidity. Across the full composition range, however, the dominant variation in the modulus follows the composition-dependent structural length scale (Fig. \ref{fig:mech}b). Within the Cauchy-Born framework, the mechanical response is therefore governed primarily by mesoscale coarsening, with local tetrahedral packing providing a smaller correction.

These results resolve the competition introduced at the beginning of this section. Increasing the large-particle fraction simultaneously enhances selected local packing motifs and coarsens the gel network. The first effect is nonmonotonic and remains localized within compact tetrahedral aggregates, whereas the second acts systematically across the full composition range and changes the dimensions of both clusters and voids. Particle size therefore sets the baseline mechanical response through the mesoscale network geometry, while packing superimposes a weaker composition-specific correction.


\section*{Conclusions}
We used hydrodynamically resolved simulations to isolate how bimodal composition changes the structure of colloidal depletion gels at fixed total particle volume fraction and a small-to-large radius ratio of 1:2. Increasing the large-particle volume fraction introduces new heterotypic angular configurations and substantially increases the formation of tetrahedral structures. However, the additional rigid motifs remain concentrated in compact local aggregates. Neither the ensemble-averaged coordination number nor the characteristic tetrahedral-aggregate size changes appreciably, indicating that enhanced local packing does not produce a more extended rigid backbone.

The dominant composition-dependent effect is instead mesoscale coarsening. The Voronoi-volume, void-size, and cluster-size distributions shift systematically as large particles replace small particles, and these changes largely collapse after normalization by a composition-dependent particle scale. The elastic modulus estimated using Cauchy–Born theory follows this effective structural length more closely than it follows tetrahedral abundance, although a smaller packing-related enhancement remains at the composition that maximizes tetrahedral participation. Thus, at the moderate size disparity considered here, particle size matters more than local packing: composition controls the predicted mechanical response primarily by setting the mesoscale network geometry, while local densification provides a secondary correction.

These findings suggest a reduced description of bimodal colloidal gels, in which composition enters first through an effective structural length scale and only second through motif-specific packing. The conclusion should not yet be assumed universal. At larger size ratios, small particles may occupy interstitial regions, alter load transmission between large particles, or form population-specific sub-networks. Changes in attraction range, size-dependent surface chemistry, or interaction non-additivity may similarly amplify the role of local packing. Determining the conditions under which a bimodal gel crosses from size-controlled to packing-controlled behavior represents a natural direction for future work.




\section*{Acknowledgements}
Financial support for this study was provided to S.J. and A.M. by the National Science Foundation (PMP-2025613) and to R.A.C. and S.J. by NASA ROSES FINESST (80NSSC23K0015). Computational resources were provided by the Massachusetts Green High-Performance Computing Center in Holyoke, MA.



\balance

\renewcommand\refname{References}

\bibliography{references} 
\bibliographystyle{rsc} 

\end{document}